\documentclass[preprint,12pt]{elsarticle}

\usepackage{epsfig}

\usepackage{amssymb}

\journal{ }

\begin{document}

\begin{frontmatter}
\title{Current-Generating Mechanism for El Ni\~{n}o \& La Ni\~{n}a, an Data Evidence from Integrated ESMD Method\footnote{I note that the ``\emph{current-generating mechanism for El Ni\~{n}o and La Ni\~{n}a}" had been finished on November, 20, 2017, and the patent application for ``\emph{Integrated ESMD Method for space-time data analysis}" had been submitted to the patent office of the People's Republic of China on January, 11, 2019.
   \textbf{Relative to the conventional empirical orthogonal function (EOF) approach, the advantage of the integrated version of our ESMD method goes without saying (please see the fresh figures in the last)!} }}


\author{Jin-Liang Wang}

\address{ {Research Institute for ESMD method and Its Applications,}\\
{College of science, Qingdao University of Technology,
Shandong, China (266520).}\\
 {E-mail: wangjinliang0811@126.com}
}
\begin{abstract}
 Different to classical theories which emphasize east--west feedback along the equator, data analyses via the extreme-point symmetric mode decomposition (ESMD) method support a new viewpoint that the El Ni\~{n}o--Southern Oscillation (ENSO) cycle is horizontally dependent, rather than vertically dependent. The consistency between the ocean current changes and sea-surface temperature anomalies results in a current-generating mechanism for El Ni\~{n}o and La Ni\~{n}a events. The corresponding new findings are as follows: (1) The appearance of the Central Equatorial Countercurrent (CECC) and the enhancement of the North Equatorial Countercurrent (NECC) are primary causes for the occurrences of positive anomalies in the central and eastern Pacific, respectively. (2) In addition to the contribution from the CECC, initiations of the eastern and central Pacific types of El Ni\~{n}o events are mainly caused by the southward transport of warm water from the high-temperature sea-area off Central America, and from the eastern low-speed gap of the NECC, respectively. (3) The northern and southern branches of the South Equatorial Current contribute to the negative anomalies in the eastern and central Pacific, respectively. Particularly, the later transfers cold water from the southeast to the central equatorial Pacific in a direct way, and this implies the crack of the ENSO-asymmetry problem.
  \end{abstract}

\end{frontmatter}



\baselineskip = 18 pt

\section{Introduction}
\setcounter{equation}{0}

El Ni\~{n}o and La Ni\~{n}a events with anomalous warm and cold sea surface temperatures (SSTs) in the tropical Pacific, respectively, are major climate phenomena that are related to extreme weather conditions worldwide (Cai et al. 2015a). The inter-annual oceanic shift between them is usually considered to be a consequence of ocean-atmosphere interactions.
  Together with the ``Southern Oscillation'' atmospheric counterpart, defined by the sea-level pressure (SLP),  this ocean-atmosphere oscillation is seen to be a self-sustaining cycle dubbed the ``El Ni\~{n}o Southern Oscillation (ENSO)''  (Enfield 1989; Neelin et al. 1998; Wang and Picaut 2004; Collins et al. 2010). In research, generally ``ENSO'' has a non-specific meaning; however, in the present article it is used to denote the aforementioned self-sustaining cycle. Relatively, the terms ``El Ni\~{n}o'' and ``La Ni\~{n}a'' are only used in a narrow sense
to indicate any SST anomalies in the equatorial Pacific, and their occurrence mechanisms are mainly concerned in the present study. To develop a better understanding of the ENSO, together with giving a review of existing classical theories, we trace back to the original problem and start the investigation from two fundamental physical facts. Other than the conventional empirical orthogonal function (EOF) method given by Lorenz (1956) for space-time data analyses, our new approach, named the extreme-point symmetric mode decomposition (ESMD) method (Wang and Li 2013, 2015),
is adopted here.

According to the locations of the SST anomalies, El Ni\~{n}o and La Ni\~{n}a are classified as the eastern Pacific (EP) type (with significant anomalies in the eastern equatorial Pacific) and the central Pacific (CP) type (with significant anomalies in the central equatorial Pacific)
(Kao and Yu 2009; Yeh et al. 2009; Capotondi et al. 2015). Historically, the EP type was recognized first, and was
seen as the canonical type. Relative to the EP type, considerably less research has been devoted to the CP type. With regards to El Ni\~{n}o, the CP type has been referred to as the ``date-line El Ni\~{n}o'' (Larkin and Harrison 2005), ``El Ni\~{n}o Modoki'' (Ashok et al. 2007) and ``warm-pool El Ni\~{n}o'' (Kug et al. 2009).

Many classical theories have been devised to explain the ENSO-cycle mechanism, such as positive Bjerknes feedback (Bjerknes 1969; Wyrtki 1975), delay oscillator theory (Suarez and Schopf 1988; Battisti and Hirst 1989) and recharge oscillator theory (Jin 1997; Jin and An 1999). As indicated by Xu et al. (2017), the EP type of El Ni\~{n}o is the conventional one, and its temporal evolution can be explained mostly by these classical theories, whereas there is still some debate about the mechanism responsible for the initiation of the CP El Ni\~{n}o. Among these mechanisms, there are three representative ones: (1) Ashok et al. (2009) argued that the same thermocline-upwelling feedback mechanism operates for the both types of El Ni\~no, except that the region of the upwelling shifts from the eastern equatorial Pacific for the EP type to the central equatorial Pacific for the CP type. (2) Kug et al. (2009) argued it is the zonal ocean advection at the equator that contributes to the development of the CP type. (3) Yu et al. (2010) suggested it is the forcing from the subtropical and extra-tropical atmosphere that excites the CP El Ni\~no.  It is indeed possible that these studies have all hinted at various truths to the actual physical mechanisms at play. In addition, the problem of ENSO asymmetry is also a controversial issue (Okumura et al. 2011; Cai et al. 2015b; Chen et al. 2016; Levine et al. 2016). An extreme El Ni\~{n}o is characterized by a disproportionately warm maximum SST anomaly in the eastern equatorial Pacific, whereas the anomaly centres of weak El Ni\~{n}o and extreme La Ni\~{n}a events
are situated in the central equatorial Pacific (the centre of a weak La Ni\~{n}a is located further towards the east) (Cai, 2015a). These imply that the occurrence mechanism of El Ni\~{n}o differs from that of La Ni\~{n}a, and it is inappropriate to describe these two events with a single dynamical model confined to the equator. Note that, since the adiabatic motion of the water plays a key role in the oceanic circulation
and climate (Huang 2015), El Ni\~{n}o and La Ni\~{n}a may be caused by different currents.

In order to get a deeper understanding on ENSO cycle, there is a necessity to restate a fundamental fact: \emph{In a self-sustaining climate system, in addition to the thermal processes related to evaporation and precipitation, there should be a dynamical ring, as depicted in \textsf{Fig.1}}. To refer to the work given by Huang (2015) again, this ring dominates the whole system. For example, during the initiation of El Ni\~{n}o, a positive SST anomaly may form in response to a change in the flow field, and its evolution should, in turn, gradually cause a change in the SLP field and a corresponding adjustment of the wind. Any objective ENSO model should follow this rule. In fact, by and large, researchers have automatically obeyed this rule. The diversity of their models lies in different hypotheses and simplifying. To keep it in mind, we re-understand the previous three classical ENSO theories as follows:

(1) In the positive Bjerknes feedback mechanism, which involves Walker circulation, the ENSO cycle is described by an SST anomaly in the EP, an east-west SST gradient and a wind anomaly in the EP. Under the hypothesis that all the adjustments are along the equator, the SST gradient results in  air-pressure differences which, in turn, affect the Walker circulation (its lower segment in the EP is actually a westward wind). The current anomaly along the equator is taken as a default response to the wind anomaly, and is hence omitted.

(2) Next is the delay oscillator model, where the sequence goes: an SST anomaly in the EP, a wind anomaly, Rossby and Kelvin waves and then an SST anomaly. Here, the effect of currents (include zonal and vertical flows) is combined with the wind, and the impact of waves (triggered by wind anomalies) to the depth adjustment of the  thermocline in the EP is stressed. \emph{There is another understanding to it: Maybe
 this adjustment is a response to current change rather than to wave stirring. In the far east Pacific, in case a cold (warm) water in the upper ocean is replaced by a warm (cold) water, it may result naturally in an SST anomaly and a thermoclinic adjustment.} In view of the fact that the upwelling region (along the Peruvian coast) is to the south of the equator,
if this idea is true, the thermoclinic adjustment at the equator should be merely a result of, rather than a cause of, an SST anomaly.
Moreover, a down-welling Kelvin wave can deepen the thermocline,
but it does not necessarily add to the thickness of the ocean surface mixed layer that determines the heat content; indeed, we should not forget that the observed sea surface height is negative for this case. Additionally, in this model, the SLP anomaly between the SST and the wind is omitted. Yet, in a common sense consideration, the response of the SLP to the SST and its effect on the wind depend on their distributions in space (maybe in the whole tropical Pacific, even in the extra-tropical Pacific), and it is unconvincing to confine them to the equator with a trivial hypothesis, as is made for the Bjerknes feedback. Details of their precise relationships are still awaiting further investigation.

(3) The recharge oscillator model can be seen as a developed version of the previous two, where emphasis is placed on thermocline feedback. The affecting sequence of this models is: an SST anomaly in the CP and EP (with a thermoclinic anomaly in the west), a wind anomaly (accompanied by an abrupt adjustment of the thermocline), a current anomaly (including zonal and meridional flows), and then an SST anomaly in the CP and EP (caused by an adjustment of the thermocline). From the viewpoint of the fundamental dynamic ring, its progress lies in stressing the effects of zonal and meridional currents.
According to the review given by Jin and An (1999), thermocline feedback is known to play two roles in the ENSO: (i) a positive reinforcement for an SST anomaly to grow, and (ii) a turn-around mechanism for its phase transition. Though this viewpoint is overwhelming, to verify or deny it is not an easy matter. What we can say is that, this knowledge origins from the hypotheses that the ENSO cycle is a self-sustaining ocean--atmosphere system confined to the equator, and the SST in the CP and EP should be adjusted by an east--west seesaw oscillation.
With the same understanding as for the previous models, not to say the SLP and wind anomalies are spatially dependent,  a change of current may directly lead to a SST anomaly. So, the growth of a positive SST anomaly may be sustained with a supply of warm water, and a phase transition may occur naturally in case the warm water is replaced by a cold water, and vice versa. If these are verified, there will be no need to consider the underlying thermocline.
To trace back for the term ``El Ni\~{n}o", it was originally used to refer to the warm current that sets
southward each year along the coast of southern Ecuador and northern Peru during the southern hemisphere summer when
the southeast trade winds are weakest (Enfield 1989).
Now that the warm current is from the north, it should pass through the equator
and contribute to the concerned anomaly of SST. Probably, this temperature-increasing intensity
can not be reached by placing hope merely on the depression of cold upwelling
with a deeper thermocline. Additionally, under the framework of this conceptual model, meridional currents that induce recharging or discharging should be symmetric with respect to the equator. However, reality does not appear to follow these assumptions and assertions. Probably, it is just the current asymmetry who results in the ENSO asymmetry. To verify this, the specific locations of them need to be clarified.

Based on the foregoing summary, we make a bold conjecture here:
\emph{the ENSO cycle is horizontally dependent rather than vertically dependent}, and any objective ENSO model should follow the full dynamic ring depicted in \textsf{Fig.1}, where its horizontal scope should include, at least, the whole tropical Pacific. The reason for this is that the northeast and southeast trade winds on both sides of the equator cover the entire tropical Pacific. Refer to the study by Yu et al. (2011), the scope of SLP anomalies may extend to the north extra-tropical Pacific.

For every two neighbouring factors among the SLP, wind, current and SST anomalies, the relationships between them  are waiting for detailed investigations. In fact, the current between the wind and the SST cannot be treated trivially. On the one hand, different ocean currents have different responses to the wind, while on the other hand, SST variations caused by heat transmission rely not only on the current strength but also on the water temperature. In the present study, we focus our attention on clarifying the relationship between changes of tropical currents and anomalies of SST, where the sensitive current passages for the occurrences of El Ni\~{n}o and La Ni\~{n}a are mainly concerned. As a default, only inter-annual time scales are considered.

In general, there are four prominent zonal surface currents in the tropic Pacific: the North Equatorial Current (NEC), North Equatorial Countercurrent (NECC), South Equatorial Current (SEC), and South Equatorial Countercurrent (SECC) (Reid 1959, Wyrtki 1974). The SEC is divided further into northern and southern branches by the equator (Wang \& Wu 2013), which we refer to as NSEC and SSEC, respectively. In addition, there is also an intermittent surface current (usually appears in the west Pacific) which flows from west to east along the equator. Delcroix et al. (1992) and Johnson et al. (2000) suggested that this current differs from the known Equatorial Undercurrent (EUC) and it is caused mainly by a westerly wind burst. So they referred to it as the ``eastward equatorial current" to clarify this difference. To disagree with them, Wang and Wu (2013) suggested that it is the surface part of the shoaled EUC. In addition, in a study of El Ni\~{n}o by Lagerloef et al. (2003) they simply called this a ``surface current." In fact, in the documented studies on ENSO, together with the westward NSEC, it is usually included in the vague calling ``zonal current" or ``zonal advection" (Collins et al. 2010). Thus, there is a necessity to give an appropriate name to this surface current regardless of its relationship with the EUC (because the word ``undercurrent" is improper for this case). Notice that all of the defined Currents flow from east to west, all of the Countercurrents flow from west to east and this one lies between the NECC and SECC, we refer to it as the ``Central Equatorial Countercurrent (CECC)."

In addition to these six zonal currents  (see \textsf{Fig.2a}), there are two meridional ones: the cold California Current and the Peru Current
from high-latitudes, which connect the NEC and NSEC \& SSEC  at the eastern boundaries of the Pacific, respectively. Particularly, the latter charges the cold-tongue region, and affects the equatorial SST significantly.
We note that the upwelling along the Peruvian coast only occurs as an accompanying water-compensation to the equator-ward flow, and the cooling effect of the Peru Current to the equator cannot be solely attributed to it.

As early as 1973, Wyrtki was aware of the connection between the transport of NECC and the occurrence of El Ni\~{n}o (the NECC carries warm water into the eastern Pacific and the fluctuation in its strength leads to temperature anomaly off Central America).
This viewpoint was very insightful, but few people have cared it for so long a time. It is a pity! Along with space-time data analyses of temperature and flow fields in the following sections, we have found that the NECC indeed plays a key part in generating El Ni\~{n}o. Besides the NECC, here all of the mentioned currents are taken into consideration.
Moreover, our research is based on another fundamental fact: \emph{In the case where a current flows from warm area A to cold area B, heat will be transferred to the latter and the corresponding temperature in B will increase.} A comprehensive consideration of the current changes under a background temperature field yields a new generation mechanism for El Ni\~{n}o and La Ni\~{n}a.

This paper is arranged as follows: The next section describes our data and method. In Section 3, the relationship between current changes and SST anomalies are investigated; meanwhile, the sensitive current passages for El Ni\~{n}o and La Ni\~{n}a are also clarified. In Sections 4 and 5, the occurrence mechanisms for El Ni\~{n}o and La Ni\~{n}a are illustrated separately. In the final section our results are summarized.

\section{Data and the ESMD Method}

The observed $2^{\circ}\times 2^{\circ}$ monthly SST data during 1950--2017 were obtained from the improved Extended Reconstructed Sea Surface Temperature version 4 from the National Climate Data Center, USA. The Ni\~{n}o-3.4 index was provided by the Climate Prediction Center. The observed $1^{\circ}\times 1^{\circ}$ 5-day absolute geostrophic velocities during 1993--2016 were obtained from the AVISO website, France.

With regards to the occurrences of El Ni\~{n}o and La Ni\~{n}a, the most pressing problem is how to correctly reflect their space-time evolution. This requires solving a difficult three-dimensional data-analysis problem where time and space possess one and two dimensions, respectively. The EOF method developed by Lorenz in 1956 has been continually employed as a default approach for this kind of problem. However, this method depends on the matrix-decomposition theory in mathematics, which can only deal with two-dimensional problems. Restricted by this, the two-dimensional plane should be considered as a one-dimensional line (i.e., the observational sites in the plane should be artificially numbered in a certain order).
Thus, the outputted spatial modes are merely static maps that do not possess evolutionary characteristics (the corresponding time coefficients only reflect relative changes in strong and weak patterns),
which cannot objectively reflect the appearance, development, propagation and extinction
of anomalous warm or cold signals in the whole sea-area,
and cannot distinctly distinguish the time-frequency multi-scale variations during the evolutionary process.
These problems are inherent and they hinder further exploration.
Therefore, it requires a method innovation.

Five years ago we developed the ESMD method for one-dimensional data analyses (Wang and Li 2013), which has advantages in trend separation, anomaly diagnosing and time-frequency analyses. Compared with the classical Fourier transform, the popular Wavelet transform, and the Hilbert--Huang transform (also known as ``empirical mode decomposition"), the ESMD method is more suitable for scientific exploration, and it has already been used in fields such as atmospheric and oceanic sciences, life sciences, informatics, mathematics, seismology, and mechanical engineering. The corresponding algorithm is provided for your reference (see the \textsf{Appendix}).

As an example, we consider the time-series of SST at grid point ($160^{\circ}$W, $0^{\circ}$). The decomposition in \textsf{Fig 3} yields a remainder R whose change-rate ($0.11^{\circ}$C per decade) agrees well with the documented global warming trend (Hansen et al. 2010). In addition to R, there are also five modes with periods $\leq 1$yr, 1--3yr, 3-7yr, 11-13yr and $\geq 18$yr, respectively. Particularly, the interannual component with period 1--3yr acts as a dominant factor. Except Mode 1 (the seasonal component), the sum of the others can be seen as the whole interannual anomaly of SST which agrees well with the known Ni\~{n}o-3.4 index (the coefficient between them is 0.87). Therefore, the ESMD method is a good choice for analysing the time series obtained from each observational site. The space--time data analysis only require an integrating process.
 Under interannual time scales, tests on SSTs and currents (shown in \textsf{Fig.2}) have shown good cooperativity of spatial patterns to all the grid points. Hence, the integration process is feasible and the integrated version of the ESMD method can be adopted as a new approach for space--time data analyses.

\section{Consistency between Current Changes and SST Anomalies}

 It is generally recognized that all of the equatorial currents, such as the NECC, NSEC and SSEC, have zonal distributions, as shown in \textsf{Figs 2a} and \textsf{2b} (only the zonal velocities are drawn), but the results in \textsf{Figs 2c} and \textsf{2d} indicate their surprising streamlines, which do not merely belong to certain currents, where they may incline, retrace, or make many-times turns throughout the whole region. These characteristics benefit the heat redistribution in the equatorial Pacific.In particular, the equatorward transport of warm water from the north and that of cold water from the south may make a great contribution to the occurrences of El Ni\~{n}o and La Ni\~{n}a events, respectively.

 The interannual evolution of SST anomaly with respect to time and longitude shown in \textsf{Fig 4b } agrees well with the documented ENSO cycle. We note that the locations of SST anomalies in this figure are very clear and the debate about ``\emph{whether the CP type of El Ni\~{n}o is part of the ENSO asymmetry or a distinct mode?}" (Cai, 2015a) should be ended.
The super-strong El Ni\~{n}o event in 1997/98 is a typical EP type. The weak events in 1994/95, 2002/03, 2004/05 and 2006/07 belong to CP type.
As for the moderate event in 2009/10 and the strong event in 2015/16,
they are best seen as mixed type ones. By the way, in case the time span for \textsf{Fig.4b}
is extended to 1950--2016, then the strong events in 1957/58, 1965/66, 1972/73 and 1986/87, and the super-strong event in 1982/83 are also included. By and large, their positive anomalies, which originated in the EP, were significant. This is why the EP El Ni\~{n}o was considered to be the conventional type. But, strictly speaking, only the event in 1982/83
(mainly dominated by an anomaly in the cold-tongue region, similar to that in 1997/98) belongs to the EP type, while the other four are best to be seen as mixed-type events since their anomalies in the CP or mid-east regions were also significant. As for La Ni\~{n}a events,
the strong ones in 1998/2000, 2007/08, 2010/11 belong to the CP type,
the weak one in 1995/96 belongs to the EP type, and the other weak one in 2005/06 is a mixed type. Before 1993 there were also many strong and weak La Ni\~{n}as. By and large, most of the strong ones, such as those in 1970/71, 1973/74, 1975/1976 and 1988/89, belong to the CP type. Almost all the weak ones belong to the EP or mixed types.

The eastward current anomaly referred to as the CECC has been proposed as being either a theoretical mechanism for El Ni\~{n}o generation (advection-reflection oscillator, Picault et al. 1996), as positive zonal-advection feedback (Jin et al. 2006 and subsequent citations), or a low-frequency component of ocean--atmosphere interaction that leads to state-dependent noise forcing (Puy et al. 2015, Levine et al. 2016). In order to verify these the evolution process of CECC is also illustrated in \textsf{Figs.4a}.
A comparison between \textsf{Figs 4a} and \textsf{4b} shows a high consistency between the occurrences of CECC and El Ni\~{n}o. As the prosperous periods concerned, the CECC leads the SST anomaly by 2--3 months which accords well with the result given by Lagerloef et al. (2003). This consistency has verified that the existence of CECC is favourable for the occurrence of El Ni\~{n}o.
In fact, it is a natural thing. The reason is that the eastward transport of warm water from the warm-pool must increase the temperature of the eastern region. However, the occurrences of El Ni\~{n}os, at least for EP types, cannot be solely attributed to the CECC. For example, for the event in 1997/98, the strongest SST anomaly occurred at the far eastern end of the Pacific where the CECC became very weak. Hence, there must be other contributors in the EP, where the most likely candidate is the NECC.

According to \textsf{Fig 4d}, in the east Pacific the equatorward transport of warm water from NECC may occur at any longitude. In particular, there are two strong intervals: one lies between $140^{\circ}$W--$110^{\circ}$W and the other lies to the east of $90^{\circ}$W. If limit them by $3^{\circ}$N and $5^{\circ}$N, then we obtain two regions for the equatorward flow. These regions are very sensitive to El Ni\~{n}o, so we refer to them as ``Ni\~{n}o-A" and ``Ni\~{n}o-B" regions, respectively (see \textsf{Fig 2c}).
It is very interesting to find that Ni\~{n}o-A region is a meridional passage which connects the east low-speed gap of the NECC (see \textsf{Figs.5a} and \textsf{6a}) and the equator.
We note that the lowest speed of this gap is near $110^{\circ}$W, which agrees well with the finding obtained by Johnson et al. (2002). It follows from \textsf{Fig 2c} that some of the NECC streamlines in Nov. 2014 took clockwise rotations from the slow-down interval $140^{\circ}$W--$110^{\circ}$W and turned back coincidentally across the Ni\~{n}o-A region. In addition, the sensitivity of the Ni\~{n}o-B region is due to the existence of a high temperature sea-area off Central America. Any equatorward transport from this area may bring warm water to the equator.

It follows from \textsf{Fig.2d} that the inclined streamlines of the SSEC dominate most southeastern sea areas. In particular, some of them thrust at the central equatorial Pacific like swords, and they can transfer cold water into this area in a direct way. The variation of ${V}_2$ in \textsf{Fig.6b} verifies this. With this understanding,  ``Ni\~{n}a-A" region is delimited. In addition, notice that the Peru current along the coast of South America always supplies cold water to the NSEC and SSEC, its strength needs to be determined. Thus, Ni\~{n}a-B region is delimited.
 We note that the selection of these four sensitive regions had involved in many trials and the results shown in \textsf{Fig 4c} were under the best choice.
 For simplicity, we refer to the averaged equatorward velocities in Ni\~{n}o-A, Ni\~{n}o-B and Ni\~{n}a-B regions as the corresponding indices separately. As for that of Ni\~{n}a-A, it has a different definition.
Since it is used to detect the inclined streams from the southeast to the central equatorial Pacific, the dominating zonal component of the velocity is mainly concerned  (see \textsf{Fig.7b}). As for the equatorward component (see \textsf{Fig.7c}), it can be borrowed to estimate  the slant angle of the mean stream, which in turn determines the anticipated scope of the negative anomaly on the equator.

In addition to the contribution of the CECC, the occurrence of positive SST anomalies should be related to the four  indices in the EP. A comparison of \textsf{Figs.4b} and \textsf{4c} together with \textsf{Fig.7b}
shows good consistency between these indices and the anomalies: (1) The prominent maximum value of Ni\~{n}o-B index in 1997 agrees well with the positive SST anomaly in the EP (the leading time was not obvious). For this case, Ni\~{n}o-A index just decreased to zero, so the anomaly in the EP was fully dominated by Ni\~{n}o-B index. (2) The prominent maximum values of Ni\~{n}o-A index  in 1994, 2002, 2004 and 2006 agree well with the positive SST anomalies in the CP (the leading times were 0--3 months). (3) Both Ni\~{n}o-A and Ni\~{n}o-B indices were high during the scenarios in 2009/10 and 2015/16. In addition, the two Ni\~{n}a indices had also taken part in
all these processes, and the intensity of its weakening affected that of El Ni\~{n}o.

In addition to clarifying the relationship between the current indices and the SST anomalies, there is a need to compare the evolutions of the zonal currents and the corresponding SSTs. The comparison between \textsf{Figs.5a} and \textsf{5b} indicates that, before the occurrences of El Ni\~{n}o events, the NECC had already strengthened (the leading times varied from several months to 2 years), and the rapid gathering of heat led to a drastic temperature increase in the sea area off Central America. This finding accords with the result given by Wyrtki (1973). Moreover, \textsf{Fig.5e} shows that the SSEC had weakened accordingly with almost the same leading time. Hence, prior to the occurrence of El Ni\~{n}o, the NECC and the SSEC (together with the NSEC) had finished adjusting in a harmonious manner, which was possibly caused by a unitary change in the wind field.

It follows from \textsf{Figs.5b, 5c} and \textsf{5d} that, in the EP, the temperature on the equator is always lower than that of the northern region that the NECC passes through, and  higher than that of the southern region where the SSEC passes through. This provides the possibility for the appearance of a positive anomaly from the north and a negative anomaly from the south. \textsf{Fig.6c} shows that, in the far east near Ni\~{n}o-B and Ni\~{n}a-B regions, the temperature difference between the north and the south is usually maintained at about $4^{\circ}$C. A strong southward flow from the high-temperature area off Central America may result in a significant anomaly in the cold-tongue region. So, the EP El Ni\~{n}o events were always very strong. In addition, no matter how the Peru current varies, its temperature is always lower than that on the equator, with a difference bigger than $1.5^{\circ}$C. This indicates that the Peru current is not a unique water-supply source to the NSEC, but instead, there should also be a meridional transport of warm water from the north, as reflected by Ni\~{n}o-B index.

It follows from \textsf{Figs.5c} and \textsf{5e} (or from \textsf{Figs.7a} and \textsf{7b}) that, before the occurrence of La Ni\~{n}a, the SSEC has already strengthened (the leading time is 5--7 months). Notice that the southeast sea area was very cold (see \textsf{Fig.5d}), and a large part of the streamlines thrust at the central equatorial Pacific (see \textsf{Fig.2d}), it was possible for the SSEC to transfer cold water to the later. In addition, due to the impact of the SECC, the temperature in the central Pacific between $4^{\circ}$S--$8^{\circ}$S is higher than that on the equator. So not all inclined streams of the SSEC contribute to the cooling of the central equatorial Pacific. The water temperature of the ones to the south of Ni\~{n}a-A region could be tremendously raised on their ways. It is why this detecting region was carefully selected. \textsf{Fig.7a} verifies that the SST on Ni\~{n}a-A region is always lower than
that in the central equatorial Pacific bounded by $4^{\circ}$S--$4^{\circ}$N from $170^{\circ}$E to $140^{\circ}$W. We note that the choice of this detecting region for CP type variation is based on the actual scope of SST anomalies,
and on the sensitive passages of ocean currents. In this regard, the conventional Ni\~{n}o-4 region (bounded by $5^{\circ}$S--$5^{\circ}$N and $160^{\circ}$E--$150^{\circ}$W) is abandoned in our study.

\section{On the Occurrence of El Ni\~{n}o}

From the previous section we see that not only the CECC, but also the two Ni\~{n}o indices, contribute to the occurrence of El Ni\~{n}o. Their impacting scopes need to be clarified. The historical El Ni\~{n}o events also need to be re-understood from this new viewpoint. In addition, to verify these new findings, the best way is to browse the space-time evolution of each scenario, which can be realized by plotting a series of spatial patterns with respect to different times.

It follows from \textsf{Fig.6d} that the temperature of Ni\~{n}o-A region ($T_1$) is always higher than that on the faced equator ($T_2$), with a difference about $1^{\circ}$C. So it is reasonable for this segment of the equator absorbing heat from the north. Yet, due to the higher temperature in CP (reflected by $T_4$),
along the westward NSEC the warm water from the east low-speed gap of NECC
can only make sense on a limited scope. The comparison between ${T}_1$ and ${T}_3$ indicates that the western boundary of this scope is about $150^{\circ}$W.

 To compare the contribution of CECC and those of Ni\~{n}o-A and Ni\~{n}o-B indexes,
it needs to stress on those seven El Ni\~{n}o events marked in \textsf{Fig.4}.
As the prosperous periods concerned , \textsf{Fig.8} shows that the easternmost locations reached by the CECC usually accord with the peak values of the SST anomaly. This relationship reflects the contribution of the CECC. Note that, since the case in 2006/07 is similar to that in 2002/03, its figure is omitted here.
Our detailed results are as follows:

(1) In \textsf{a} and \textsf{c}, each line for SST anomaly (SSTA) only possesses a unique peak. This indicates that the events in 1994/95 and 2002/03 are mainly dominated by the CECC.

(2) For the double-peak case in \textsf{d} and \textsf{e}, those to the east of 210 (that is $150^{\circ}$W) should ascribe to the effects of Ni\~{n}o indices. Recalling \textsf{Fig.4c}, we see that both Ni\~{n}o-A index and the CECC are the dominant factors to the event in 2004/05. As for the 2009/10 event, the effect of Ni\~{n}o-B index should be also included, since the SSTA has also a potential peak to the east of 250 (that is $110^{\circ}$W).

(3) For the case in \textsf{b}, the unique peak is at the far eastern end. This indicates that, for the event in 1997/98, the positive anomaly in the EP was mainly dominated by Ni\~{n}o-B index. Yet the contribution of CECC cannot go unnoticed: its intensity was so tremendous that the SSEC streams traversing Ni\~{n}a-A region were forced to change their direction from northwestward to northeastward (see \textsf{Fig.7}).
To recall \textsf{Fig.4c} we see that for this case, the southward transport from the high-temperature sea area off Central America was so strong that the cold-water supply from the south was completely held back.
On this occasion, the NSEC (which was to the north of the CECC in the EP) only gained water supply from the north. Though its intensity had considerably weakened, the westward transport of warm water benefited the
development of the positive anomaly. Therefore, the east and central parts of the tropical Pacific were full of warm water, and the occurrence of this super-strong event was not at all surprising.

(4) It follows from \textsf{f} that, by and large, the SSTA for the event in 2015/16 had three peaks. The ones near 200 and 250 (that is, $160^{\circ}$W and $110^{\circ}$W) should be ascribed to the CECC
and Ni\~{n}o-B index, respectively. By the way, due to the impact of the NSEC, the peak location of the latter possibly shifted from the far eastern end. The left one near $140^{\circ}$W should be ascribed to the combined actions of Ni\~{n}o-A index and the CECC. Though Ni\~{n}o-B index for this event was as high as that in 1997/98, its intensity was lower than the later. The reason for this is that the cold-water supply from the south was not completely held back, and the subsequent transport via the NSEC and SSEC weakened the intensity of the positive anomaly.  Certainly, the relatively weak CECC was also a major cause.

Before analyzing the space-time evolution characteristics of the El Ni\~{n}o events, there is a necessity to understand the phenomenon shown in \textsf{Fig.2e}:  \emph{i.e. the positive anomaly in the CP (near the data-line) and that to the northeast seem to be connected in Nov. 2014}. Indeed, this kind of pattern accords with that given by Yu et al (2010, 2011), who argued that this type of anomaly spreads from the northeast, and the occurrence of the CP El Ni\~{n}o was due to subtropical forcing through this passage. The corresponding streamlines (which belong to the NEC) in our \textsf{Fig.2c} confirm
that the anomaly surely spreads from the northeast.
However, quite unexpectedly, these streamlines do not intrude into the CP in a direct way! Instead, they make sharp turns to the east near $8^{\circ}$N (which are merged in the NECC), and then turn back from the low-speed eastern gap of the NECC into the NSEC.

Moreover, we have an additional finding: \emph{the counter-clockwise rotation of the streamlines in \textsf{Figs.2c} and \textsf{2d} indicate the existence of an eddy with its centre near ($140^{\circ}$W, $10^{\circ}$N), which facilitates the exchange  between the equatorial Pacific and the north-eastern subtropical Pacific.} Due to a lack of knowledge, we cannot confirm if this eddy has been documented. Since its shape is like a big triangle, for convenience we will call the ``tropical Pacific triangle (TPT) eddy''. Notice that the right hand side of this eddy is a high-temperature sea area, and any transport of warm water from this area must lead to a positive anomaly, as shown in \textsf{Fig.2e}.

\emph{Hence, can the anomalous signals from the northeast for the other cases intrude into the CP in a direct way?} \textsf{Fig.6b} gives a negative answer. Before the occurrences of El Ni\~{n}o, the developments of the CECC always impacts the date-line region, and results in northward flows (reflected by ${V}_1$). So, intrusion from the north is impossible, and the positive anomaly in the CP should be solely ascribed to the CECC.
By the way, to the north and south of the equator, the impacted flows are almost symmetric with respect to time. But, on the whole, the northward velocity is relatively higher. This should be ascribed to the impact of the SSEC from the southeast.

A series of spatial patterns with respect to different times in \textsf{Fig.9} has reappeared the space-time evolution of
the scenario during 1997--1998. They can be used to check the previous judgments. From the first sub-figure, we see the phenomenon in November 2014
also occurred in January 1997. So, the positive anomalies near the date-line should be also understood as being a result of the CECC. The subsequent evolution in March--May 1997 showed clearly that these anomalies extend to the east along the CECC. It is worth mentioning that the extending speed is much quicker than the flow velocity of the CECC. This is because, there exists a west--east temperature gradient, and any eastward flow at any longitude must result in a positive anomaly to its downstream.
This differs from the case of transferring an object from site A to site B by a flow. From \textsf{Fig.8b} we see the CECC covered all the equatorial region from $160^{\circ}$E to $100^{\circ}$W, and it is not strange for this quick extending. In addition to the positive anomalies in the CP, there were also some present along the coast of South America from $30^{\circ}$S to $0^{\circ}$. In the previous section, we have stressed on the southward intruding from the high-temperature sea-area
off Central America. But the corresponding streamlines indicated that the southernmost latitude of this intrusion was only about $6^{\circ}$S (the figure is omitted), and there should be other reasons for the anomaly seen between $6^{\circ}$S--$30^{\circ}$S. From the previous analysis we see, due to the impact of the CECC, the SSEC streams across Ni\~{n}a-A region were forced to flow to the east. They took a clockwise rotation near $100^{\circ}$W where they met the intruding flow from the north. Subsequently, some of the SSEC streams from the south near this longitude were also forced to take a clockwise rotation, which transferred relatively warm water to the coast of South America (who possesses the lowest temperature all the year round). This may be the reason for the occurrence of this off-equatorial anomaly.
In addition, during the decay process it  reflected clearly an eastward withdrawing characteristic. This should mainly be ascribed to the disappearance of the CECC. Meanwhile, the intensity of the anomaly
also gradually weakened, which could have been a response to the decreasing of Ni\~{n}o-B index.

It follows from \textsf{Fig.2e} that, in November 2014, just before the arrival of the 2015/16 El Ni\~{n}o, the presence of strong flows from the Ni\~{n}o-A and B regions together with a strong CECC from the warm-pool generated a favourable situation in the mid-eastern Pacific. It is expected that the negative-anomaly region would be occupied by warm water from both sides. The space-time evolution in \textsf{Fig.10} verifies this.
Particularly, the impact of Ni\~{n}o-A index in the initiating stage of this event was very typical, and the CP type ones in 2004/05 and 2009/10 also underwent the same evolutionary process. \textsf{Fig.10} also shows a distinct difference between the anomaly near the date-line and that from the northeast. Furthermore, as analysed in the previous, the presence of a prominent pattern in the northeast was mainly due to the transport by the TPT eddy from the high-temperature sea area off Central America. The start time of this transport can be dated back to January 2014
when the equatorial region was still charged by a negative anomaly. In addition, there was no significant anomaly in the far eastern region, though Ni\~{n}o-B index was very high for this case. Recalling \textsf{Fig.2c}, we see that the reason for this is that the transport of cold water from the south still existed for this case.

\section{On the Occurrence of La Ni\~{n}a}

The occurrence of La Ni\~{n}a in the equatorial region relies on the supply of cold water from the southeast by the NSEC and SSEC. In the case where the NSEC dominates, it will be an EP La Ni\~{n}a, and on the contrary it will be a CP La Ni\~{n}a.

Under normal circumstances, the NSEC is composed of warm water from the north and cold water from the south. Particularly, in the case where the northward branch of the Peru current becomes stronger, cold water in the NSEC dominates, and a negative anomaly will first appear in the far eastern end. The subsequent development will result in an EP La Ni\~{n}a, such as that in 1995/96.

 \emph{Why is the EP type of La Ni\~{n}as so weak?} The primary reason is that the far eastern end of the equator includes a cold-tongue region, and the temperature difference between the southern sea-area (from $4^{\circ}$S to $8^{\circ}$S) and it is no more than $2^{\circ}$C, no matter how the Peru current varies.  \textsf{Fig.6c} shows that the cooling in 1995/96 is only about $1^{\circ}$C relative to normal conditions. Relatively speaking, the temperature difference between the cold-tongue region and the central equatorial Pacific is much higher. So, CP La Ni\~{n}a events are usually very strong.

 The explanation of the strong La Ni\~{n}a events in 1998/2000, 2007/08, and 2010/11 is very simple because the dominant factor was just the Ni\~{n}a-A index. In the case where the westward branch of the Peru current becomes stronger, it provides enough cold water to the SSEC. The subsequent transport by the SSEC leads to a negative anomaly in the CP.  Certainly, in most cases the SSEC and NSEC enhance simultaneously, which is possibly caused by an enhancement of the southeast monsoon.  So during the course of strong events, the enhancement of the NSEC is also beneficial.  But, as the negative anomaly concerned, the impact of it is much lower
 that of SSEC.

\emph{Why is the phase transition from El Ni\~{n}o to La Ni\~{n}a so rapid?} It follows from \textsf{Fig.2f} and \textsf{Fig.9} that, as the positive anomaly withdraws to the east, a cold centre appears in the central equatorial Pacific. The latent reason for this exists in the disappearance of the CECC. \textsf{Fig.6b} shows that, along with the vanishing CECC,
the SSEC was released and it rapidly controlled the date-line region.
Moreover, the enhancement of the SSEC and NSEC also created favourable conditions for the development of a negative anomaly. Thus a strong La Ni\~{n}a was expected subsequently. Relatively speaking, the phase transition from La Ni\~{n}a to El Ni\~{n}o needs more time. In view of the dynamic ring in \textsf{Fig.1}, adjustments in the SLP field, the wind field, the current field and then the temperature field, are needed.

\section{Conclusions }

It is a fundamental fact that the wind, ocean current, SST, and SLP form a dynamic ring where they act in a fixed order. To start from this fact, we have reviewed the three classical theories for ENSO, and developed our understanding whereby \emph{the ENSO cycle is horizontally dependent rather than vertically dependent}. The consistency between the ocean-current changes and the SST anomalies results in a current-generating mechanism for El Ni\~{n}o and La Ni\~{n}a events. Our findings are as follows:

(1) We verified that the appearance of the eastward CECC from the warm-pool is a primary cause for the appearance of the positive anomalies in the central Pacific.

(2) We found that an enhancement of the NECC makes great contribution to the positive anomaly
in the eastern Pacific.
Particularly, Ni\~{n}o-A and -B indices were defined to allow us to detect any equatorward transport of warm water from the east low-speed gap of the NECC and from the high-temperature sea-area off Central America, respectively. The detailed proofs indicate that, in addition to the contribution from the CECC, the CP and EP types of El Ni\~{n}o events were mainly dominated by Ni\~{n}o A and B indices, respectively. In addition, the happened EP type of El Ni\~{n}os were always much stronger than the CP type ones. The reason for this is that the temperature difference between the sea area off Central America to the north
and the cold-tongue region to the south was much bigger than that of the later.  Limited by the temperature gradient along the equator from the west to east, the warm stream from Ni\~{n}o-A region (which merges within the westward NSEC) only contributes to the positive anomaly to about $150^{\circ}$W.
So, for the CP type El Ni\~{n}o, in most cases the CECC is the dominant factor.

(3) We found that not only the enhancement of the NSEC, but also that of the SSEC makes great contribution to the negative anomalies on the equator.
The difference lies in the impacting regions. The form and the later take  effect in the eastern and central Pacific, respectively. Particularly,
the initiation of strong CP La Ni\~{n}a events is due to the enhancement of the SSEC which transfers cold water from the southeast to the central equatorial Pacific in a direct way. This has cracked the ENSO asymmetry problem.
In detail, Ni\~{n}a-A and -B indices were defined to allow us to detect any cold water being transferred to the central equatorial Pacific and to the far eastern end of the equator, respectively. The CP type of La Ni\~{n}as simply depends on the intensity of Ni\~{n}a-A index, yet the EP type ones rely on the game-playing between Ni\~{n}a-B and Ni\~{n}o-B indices. Only in case the cold-water supply to the NSEC from the south prevails, will a negative anomaly occur in the EP. Due to the small temperature difference
 between the equator and the sea area to the south (they are all included in the cold-tongue region), the happened EP La Ni\~{n}as were always very weak.

In addition, along with the vanishing of the CECC, the SSEC is released and it rapidly controls the date-line region. So in most cases El Ni\~{n}o events are closely followed by strong La Ni\~{n}a events. However, the phase transition from La Ni\~{n}a to El Ni\~{n}o needs more time due to the slow adjustments from the SLP, the wind to the current, and these processes contain the remaining secrets of the ENSO cycle which need further research.

\vskip 0.5cm
\noindent\textbf{Appendix: The Algorithm for ESMD Method}

Here only one-dimensional signals are considered.
It is a common sense that the local maxima and minima points are septal
with counting all the adjacent equal extreme points as one.
The decomposition algorithm is as follows (Wang and Li 2013, 2015):

\vskip 1mm\noindent\textbf{Step 1}: Find all the local extreme points (maxima points plus minima points) of the data $Y$ and
numerate them by $E_i$ with $1\leq i\leq n$.

\vskip 1mm
\noindent\textbf{Step 2}: Connect all the adjacent $E_i$ with line segments and
mark their midpoints by $F_i$ with $1\leq i\leq n-1$.

\vskip 1mm
\noindent\textbf{Step 3}: Add a left and a right boundary midpoints $F_0$ and $F_n$ through a certain approach.

\vskip 1mm
\noindent\textbf{Step 4}: Construct $p$ interpolating curves $L_1,\cdots, L_p$ ($p\geq 1$) with all these $n+1$ midpoints
and calculate their mean value by $L^*=(L_1+\cdots+L_p)/p$.

\vskip 1mm
\noindent\textbf{Step 5}: Repeat the above four steps on $Y-L^*$ until $|L^*|\leq \varepsilon$
($\varepsilon$ is a permitted error) or the sifting times attain a preset maximum number $K$. At this time, we get
the first mode $M_1$.

\vskip 1mm
\noindent\textbf{Step 6}: Repeat the above five steps on the residual $Y-M_1$ and get $M_2, M_3\cdots$ until
the last residual $R$ with no more than a certain number of extreme points.

\vskip 1mm
\noindent\textbf{Step 7}: Change the maximum number $K$ on a finite integer interval $[K_{{min}}, K_{{max}}]$
and repeat the above six steps.
Then calculate the variance $\sigma^2$ of $Y-R$ and plot a figure with $\sigma/\sigma_0$ and $K$,
here $\sigma_0$ is the standard deviation of $Y$.

\vskip 1mm
\noindent\textbf{Step 8}: Find the number $K_0$ which accords with minimum  $\sigma/\sigma_0$
 on $[K_{{min}}, K_{{max}}]$. Then use this $K_0$
to repeat the previous six steps and output the whole modes. At this time, the last residual $R$ is actually an optimal
 adaptive global mean (AGM) curve.

\vskip 1mm
 According to the fourth step, we classify the ESMD into ESMD\_I, ESMD\_II, ESMD\_III, $\cdots$.
  ESMD\_I does the sifting process by using only $1$ curve interpolated
by all the midpoints; ESMD\_II does the sifting process by using $2$ curves interpolated
by the odd and even midpoints, respectively; ESMD\_III does the sifting process by using $3$ curves interpolated
by the midpoints numerated by $3k+1$, $3k+2$ and $3(k+1)$ ($k=0,1,\cdots$), respectively.
Certainly, one can also define other schemes with more interpolating curves according to this method.
We note that ESMD\_II is superior to
ESMD\_I and ESMD\_III and it is usually taken as the default one.

\vskip 1mm Denote the data and the AGM curve by $Y=\{y_i \}_{i=1}^N$
and $R=\{r_i \}_{i=1}^N$ separately,
the variances $\sigma_0$ and $\sigma$ are defined relative to the
total mean $\overline{Y}=\sum_{i=1}^N y_i/N$ and AGM curve as follows:
\begin{equation}
\sigma_0^2=\frac{1}{N}\sum\limits_{i=1}^N (y_i-\overline{Y})^2,
\quad \sigma^2=\frac{1}{N}\sum\limits_{i=1}^N (y_i-r_i)^2.\nonumber
\end{equation}
In the applications, we usually choose $\varepsilon=0.001\sigma_0$ and
use the ratio $\nu=\sigma/\sigma_0$ to reflect the degree of optimization
for the AGM relative to the commonly used total mean.

\vskip 0.5cm
\noindent\textbf{References}

\bibliographystyle{ametsoc2014}
\bibliography{references}

Ashok, K., S. K. Behera, S. A. Rao, H. Weng, T. Yamagata, 2007: El Ni\~{n}o Modoki and its possible teleconnection.
\emph{Journal of Geophysical Research Oceans}, \textbf{112} (C11), https: //doi.org /10. 1029 /2006JC003798.

Ashok, K., and T. Yamagata, 2009: Climate change: The El Ni\~{n}o with a difference.
\emph{Nature}, \textbf{461} (7263), 481--4, DOI: 10.1038 /461481a.

Battisti, D. S., and A. C. Hirst, 1989: Interannual variability in a tropical atmosphere-ocean model: Influence of the basic state, ocean geometry and nonlinearity. \emph{J. Atmos. Sci}, \textbf{46} (12), 1687--1712.
DOI: 10.1175 /1520-0469(1989) 046$<$1687:IVIATA$>$ 2.0.CO;2.

Bjerknes, J, 1969: Atmospheric teleconnections from the equatorial Pacific. \emph{Mon. Wea. Rev.}, \textbf{97} (3),
DOI: 10.1175 /1520-0493(1969) 097$<$0163:ATFTEP$>$ 2.3.CO;2.

 Cai, W. J., and Coauthors, 2015a: ENSO and greenhouse warming. \textit{Nature Clim. Change}, \textbf{5}, 849--859, DOI: 10.1038 /NCLIMATE2743.

 Cai, W. J., and Coauthors, 2015b: Increased frequency of extreme La Ni\~{n}a events under greenhouse warming. \textit{Nature Clim. Change}, \textbf{5}, 132--137, DOI: 10.1038 /nclimate2492.

 Capotondi, A., and Coauthors, 2015: Understanding ENSO diversity. \textit{Bull. Am. Meteorol. Soc.}, \textbf{96}, 921--938,
 DOI: 10.1175/BAMS-D-13-00117.1.

 Chen, M., T. Li, X. Shen, B. Wu, 2016: Relative roles of dynamic and thermodynamic processes in causing evolution asymmetry between El Ni\~{n}o and La Ni\~{n}a. \textit{J. Climate}, \textbf{29}, 2201--2220.
 DOI: 10.1175 /JCLI-D-15-0547.1.

 Collins, M., and Coauthors, 2010: The impact of global warming on the tropical Pacific ocean and El Ni\~{n}o. \textit{Nature Geosci.}, \textbf{3} (6), 391--397, DOI: 10.1038 /ngeo868.

 Delcroix, T., G. Eldin, M. H. Radenac, J. Toole, E. Firing, 1992: Variation of the western equatorial Pacific Ocean, 1986¨C1988. \textit{Journal of Geophysical Research Oceans}, \textbf{97} (C4), 5423--5445,
 DOI: 10.1029 /92JC00127.

 Enfield, D. B., 1989: El Ni\~{n}o, past and present. \textit{Reviews of Geophysics}, \textbf{27} (1) 159--187.

 Hansen, J., R. Ruedy, M. Sato, K. Lo, 2010: Global surface temperature change. \textit{Reviews of Geophysics}, \textbf{48}, RG4004, DOI: 10.1029 /2010RG000345.

Huang, R. X., 2015: Heaving modes in the world oceans. \emph{Climate Dynamics}, \textbf{45} (11), 3563--3591,
DOI: 10.1007 /s00382-015-2557-6.

Jin, F. F., 1997: An equatorial ocean recharge paradigm for ENSO. Part I: conceptual model.
\emph{J. Atmos. Sci.}, \textbf{54} (7): 811--829, https: //doi.org /10.1175/1520-0469(1997)054 $<$0811:AEORPF$>$ 2.0.CO;2.

Jin, F. F., and S. I. An, 1999: Thermocline and zonal advective feedbacks within the equatorial ocean recharge oscillator model for ENSO. \emph{Geophysical Research Letters}, \textbf{26} (19): 2989¨C-2992, DOI: 10.1029 /1999GL002297.

Jin, F. F., S. T. Kim, L. Bejarano, 2006: A coupled-stability index for ENSO.
\emph{Geophysical Research Letters}, \textbf{33}(23), 265--288, DOI: 10.1029 /2006GL027221.

Johnson, G. C., M. J. Mcphaden, G. D. Rowe, K. E. Mctaggart, 2000: Upper equatorial Pacific Ocean current and salinity variability during the 1996--1998 El Ni\~{n}o--La Ni\~{n}a cycle. \textit{Journal of Geophysical Research Oceans}, \textbf{105} (C1), 1037--1053, DOI: 10.1029 /1999JC900280.

Johnson, G. C., B. M. Sloyan, W. S. Kessler, K. E. Mctaggart, 2002: Direct measurements of upper ocean currents and water properties across the tropical Pacific during the 1990s. \textit{Progress in Oceanography}, \textbf{52} (1), 31--61.
DOI: 10.1016 /S0079-6611(02) 00021-6.

Kao, H. Y., and J. Y. Yu, 2009: Contrasting eastern-Pacific and central-Pacific types of ENSO.
\textit{J. Climate}, \textbf{22} (3): 615--632, DOI: 10.1175 /2008JCLI2309.1.

Kug, J. S., F. F. Jin, S. I. An, 2009: Two-types of El Ni\~{n}o events: Cold tongue El Nino and warm pool El Ni\~{n}o.
\textit{J. Climate}, \textbf{22} (6): 1499--1515, DOI: 10.1175 /2008JCLI2624.1.

 Lagerloef, G. S. E., and Coauthors, 2003: El Ni\~{n}o Tropical Pacific Ocean surface current and temperature
evolution in 2002 and outlook for early 2003. \textit{Geophysical Research Letters}, \textbf{30} (10), 1514,
DOI: 10.1029 /2003GL017096.

Larkin, N. K., and D. E. Harrison, 2005: Global seasonal temperature and precipitation anomalies during El Ni\~{n}o autumn and winter. \emph{Geophysical Research Letters}, \textbf{32} (16), L16705, DOI: 10.1029 /2005GL022860.

Levine, A., F. F. Jin,  M. J. McPhaden, 2016: Extreme noise-extreme El Ni\~{n}o: how state-dependent noise forcing creates
 El Ni\~{n}o--La Ni\~{n}a asymmetry. \textit{J. Climate}, \textbf{29}, 5483--5499, DOI: 10.1175 /JCLI-D-16-0091.1.

Lorenz, E. N., 1956: \textit{Empirical Orthogonal Functions and Statistical Weather Prediction}. Statistical Forecasting Project Report 1, MIT Department of Meteorology, Cambridge, 49 pp.

Neelin, J. D., D. S. Battisti, A. C. Hirst, F. F. Jin, Y. Wakata, T. Yamagata, S. E. Zebiak, 1998: ENSO theory.
\emph{Journal of Geophysical Research Oceans}, \textbf{103} (C7): 14261--14290, DOI: 10.1029 /97JC03424.

Okumura, Y. M.,  M. Ohba, C. Deser, H. Ueda, 2011: A proposed mechanism for the asymmetric duration of El Ni\~{n}o and La Ni\~{n}a. \textit{J. Climate}, \textbf{24}, 3822--3829, DOI: 10.1175 /2011JCLI3999.1.

Picaut, J.,  M. Ioualalen, C. Menkes, T. Delcroix, M. J. McPhaden, 2009: Mechanism of the zonal displacements of the Pacific warm pool: Implications for ENSO. \emph{Nature}, \textbf{461} (7263), 481--484, DOI: 10.1038 /461481a.

 Puy, M., J. Vialard, M. Lengaigne, E. Guilyardi, 2015: Modulation of equatorial Pacific westerly/easterly wind events by the Madden-Julian oscillation and convectively-coupled Rossby waves. \emph{Climate Dynamics}, \textbf{46} (7-8), 2155-2178,
DOI: 10.1007 /s00382-015-2695-x.

 Reid, J. L., 1959: Evidence of a south equatorial countercurrent in the Pacific Ocean.
\textit{Nature}, \textbf{184} (4681), 209--210, DOI: 10.1038 /184209a0.

 Suarez, M. J., and P. S. Schopf, 1988: A delayed action oscillator for ENSO.
 \emph{J. Atmos. Sci}, \textbf{45} (21), 3283-3287.

 Wang, J. L., and Z. J. Li, 2013: Extreme-point symmetric mode decomposition method for data analysis. \textit{Advances in Adaptive Data Analysis}, \textbf{5}(3), 1350015, https:// doi.org /10.1142 /S1793536913500155.

 Wang, J. L., and Z. J. Li, 2015: \textit{Extreme-Point Symmetric Mode Decomposition Method: A New Approach for Data Analysis and Science Exploration}. Beijing: Higher Education Press (in Chinese).

 Wang, C. Z., and J. Picaut, 2004: Understanding ENSO physics--A review. \emph{American Geophysical Union},
 \textbf{147}, 21--48, DOI: 10.1029/147GM02.

  Wang, L. C., and C. R. Wu, 2013: Contrasting the flow patterns in the equatorial Pacific between two types of El Ni\~{n}o.
\textit{Atmosphere--Ocean}, \textbf{51} (1), 60--74, DOI: 10.1080/07055900.2012.744294.

 Wyrtki, K., 1973: Teleconnections in the equatorial Pacific ocean. \textit{Science}. \textbf{180}, 66--68.

 Wyrtki, K., 1974: Equatorial currents in the Pacific 1950 to 1970 and their relations to the trade winds.
\textit{Journal of Physical Oceanography}, \textbf{4}, 372--380.

Wyrtki, K., 1975: El Ni\~{n}o--the dynamic response of the equatorial Pacific ocean to atmospheric forcing.
\emph{J. Phys. Oceanogr.}, \textbf{5}, 572--584.

 Xu, K., C. Y. Tam, C. W. Zhu, B. Q. Liu, W. Q. Wang, 2017: CMIP5 projections of two types of El Ni\~{n}o and their related tropical precipitation in the twenty--first century. \textit{J. Climate}, \textbf{30} (3), 849--864, DOI: 10.1175/JCLI-D-16-0413.1.

 Yeh, S. W., J. S. Kug, B. Dewitte, M. H. Kwon, B. P. Kirtman, F. F. Jin, 2009: El Ni\~{n}o in a changing climate. \textit{Nature}, \textbf{461}, 511--515, DOI:10.1038/nature08316.

  Yu, J. Y., H. Y. Kao, T. Lee, 2010: Subtropics-related interannual sea surface temperature variability in the central equatorial Pacific. \textit{J. Climate}, \textbf{23} (11), 2869--2884, DOI: 10.1175 /2010 JCLI 3171.1.

Yu, J. Y. and S. T. Kim, 2011: Relationships between extratropical sea level pressure variations and the central Pacific and eastern Pacific types of ENSO. \textit{J. Climate}, \textbf{24} (3), 708--720, DOI: 10.1175 /2010 JCLI 3688.1.

\begin{figure}[h]
\centerline{\includegraphics[width=\textwidth]{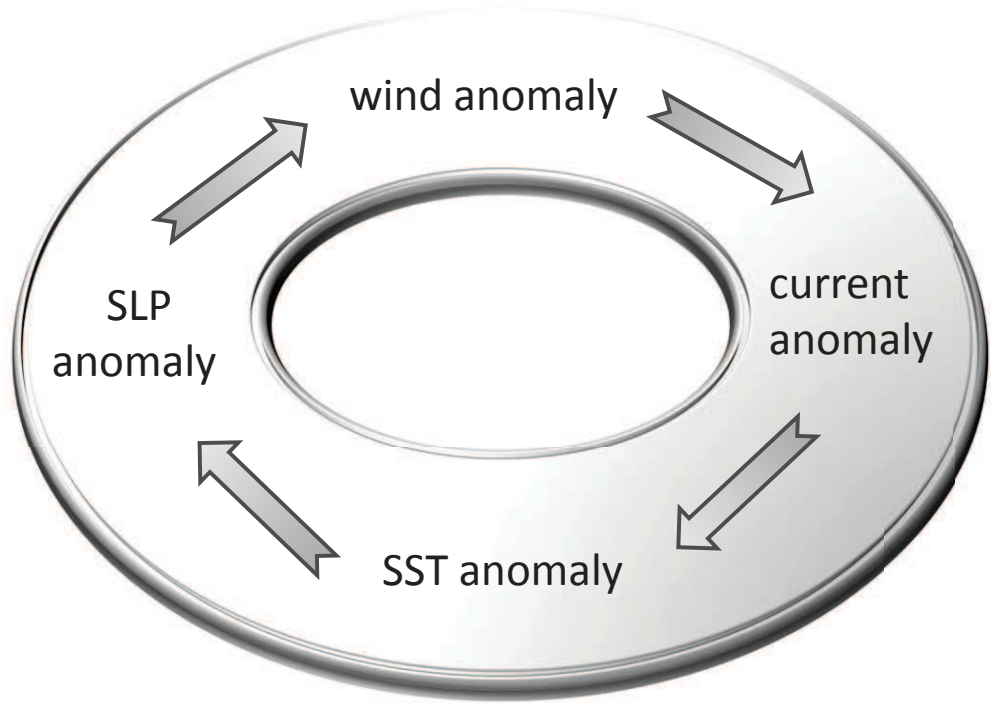}}

\caption{Depiction of the fundamental dynamical ring for a self-sustaining ocean--atmosphere climate system. }
\end{figure}

\begin{figure}[h]
\centerline{\includegraphics[width=5.1in]{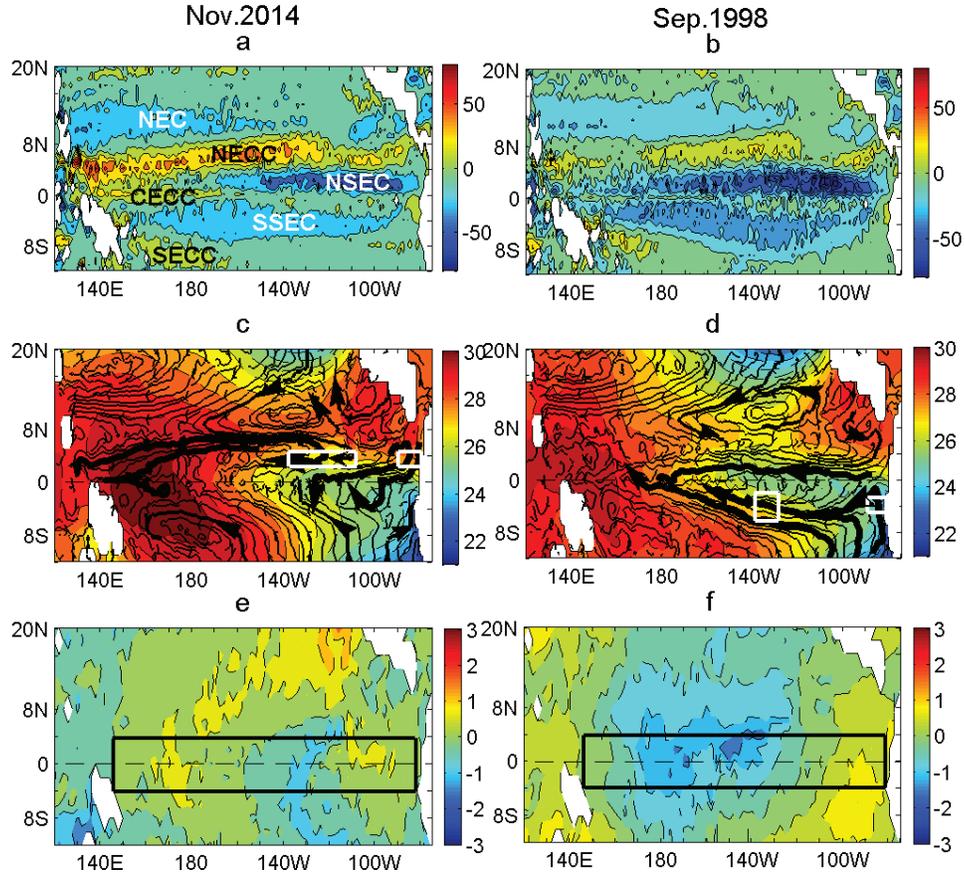}}

\caption{Comparisons of the flow and temperature fields in November 2014 (\textsf{a, c, e}) and
September 1998 (\textsf{b, d, f}). \textsf{a \& b}: Spatial locations and strengths of the equatorial currents
with interannual time scales (only the zonal velocities are drawn). The regions with white borders
denote the east low-speed gap of the NECC bounded by
$5^{\circ}$N--$7^{\circ}$N from $90^{\circ}$W to $140^{\circ}$W.
The unit for the colored bars is cm/s (positive to the east).
\textsf{c \& d}: Interannual streamlines and background temperature fields (each grid point matches with a
 certain value in the long-term SST trend).
  The two white bordered areas in \textsf{c} are the Ni\~{n}o-A and Ni\~{n}o-B regions bounded by
$3^{\circ}$N--$5^{\circ}$N from $110^{\circ}$W to $140^{\circ}$W and from $77^{\circ}$W to $90^{\circ}$W, respectively.
 The two white bordered areas in \textsf{d} are the Ni\~{n}a-A and Ni\~{n}a-B regions bounded by
$2^{\circ}$S--$6^{\circ}$S from $140^{\circ}$W to $130^{\circ}$W, and
bounded by $3^{\circ}$S--$5^{\circ}$S from $81^{\circ}$W to $90^{\circ}$W, respectively.
\textsf{e \& f}: Spatial patterns of the SST anomalies with respect to the interannual changes.
  The areas with black borders denote the region for the ENSO SST
  bounded by $4^{\circ}$S--$4^{\circ}$N from $146^{\circ}$E to $82^{\circ}$W.
  The colored bars in \textsf{c--f} have the same unit $^{\circ}$C.}
\end{figure}

\begin{figure}[h]
\centerline{\includegraphics[width=\textwidth]{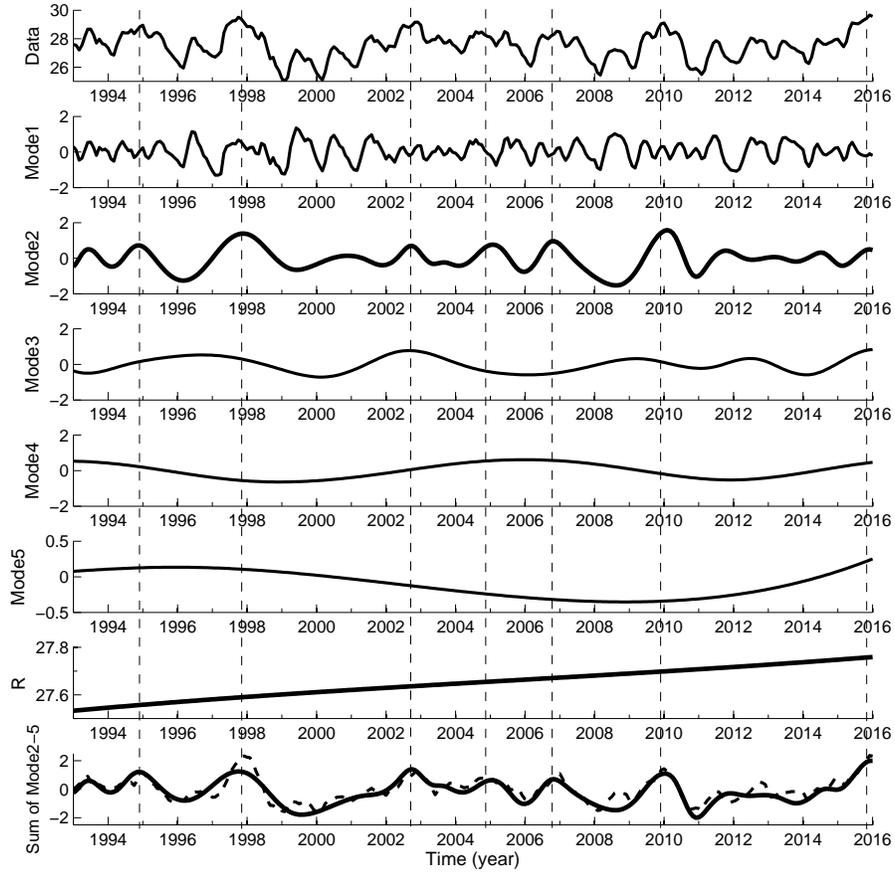}}

\caption{Decomposition results obtained using the ESMD method for the time series at grid point
 ($160^{\circ}$W, $0^{\circ}$). ``Data" denotes the raw SST data. Modes 1--5 are the decomposed modes.
``R" is the remainder mode and it reflects the effect of global warming.
The last sub-figure compares the sum of Modes 2--5 (solid curve)
 and the known Ni\~{n}o-3.4 index (the average SST anomaly in the region bounded by $5^{\circ}$S--$5^{\circ}$N from $170^{\circ}$W to $120^{\circ}$W, which is denoted by a dashed curve).
 All of the vertical axes use the same unit $^{\circ}$C.}
\end{figure}

\begin{figure}[h]
\centerline{\includegraphics[width=5.1in]{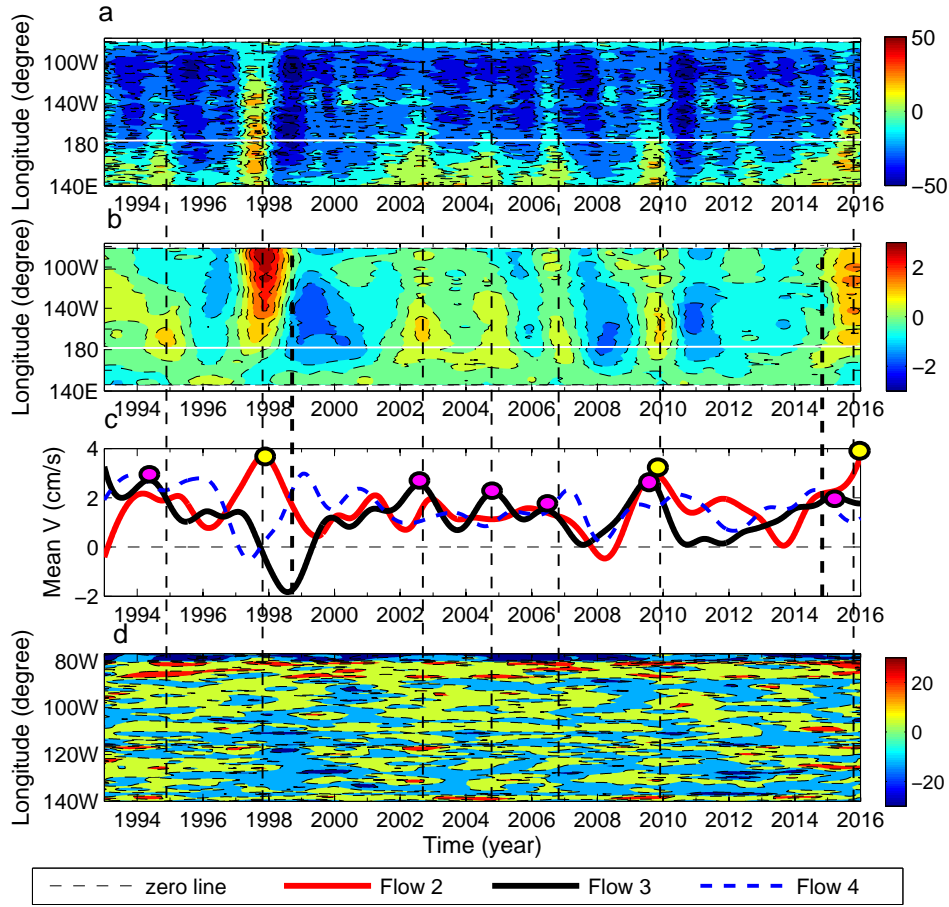}}

\caption{Consistency between the current changes and SST anomalies.
\textsf{a}: Interannual evolution of the CECC with respect to time and longitude
 (only the zonal velocity U is drawn and on each longitude, the value denotes
 the average over $2^{\circ}$S--$2^{\circ}$N). The unit for the colored bar is cm/s (positive to the east).
\textsf{b}: Interannual evolution of the SST anomaly
averaged over $4^{\circ}$S--$4^{\circ}$N, as in \textsf{a}. The unit for the colored bar is $^{\circ}$C.
\textsf{c}: Flows 1--4 are the mean equatorward velocities in the Ni\~{n}a-A, Ni\~{n}o-B,
Ni\~{n}o-A, and Ni\~{n}a-B regions, respectively (positive to the equator). The maximum points
marked by yellow circles on Flow 2 and the pink circles on Flow 3 match with the EP and CP types of El Ni\~{n}o events, respectively.
The high values marked by rectangles on Flow 1 match with the three distinct La Ni\~{n}a events.
\textsf{d}: Interannual evolution of the equatorward flow from the north
 (only the meridional velocity V is drawn, where the average is taken over $3^{\circ}$N--$5^{\circ}$N).
 The unit for the colored bar is cm/s (positive to the south).
The vertical thin dashed lines denote the prosperous periods for seven El Ni\~{n}o events.
The bold dashed lines denote the two periods considered, i.e., September 1998 and November 2014.
  }
\end{figure}

\begin{figure}[h]
\centerline{\includegraphics[width=\textwidth]{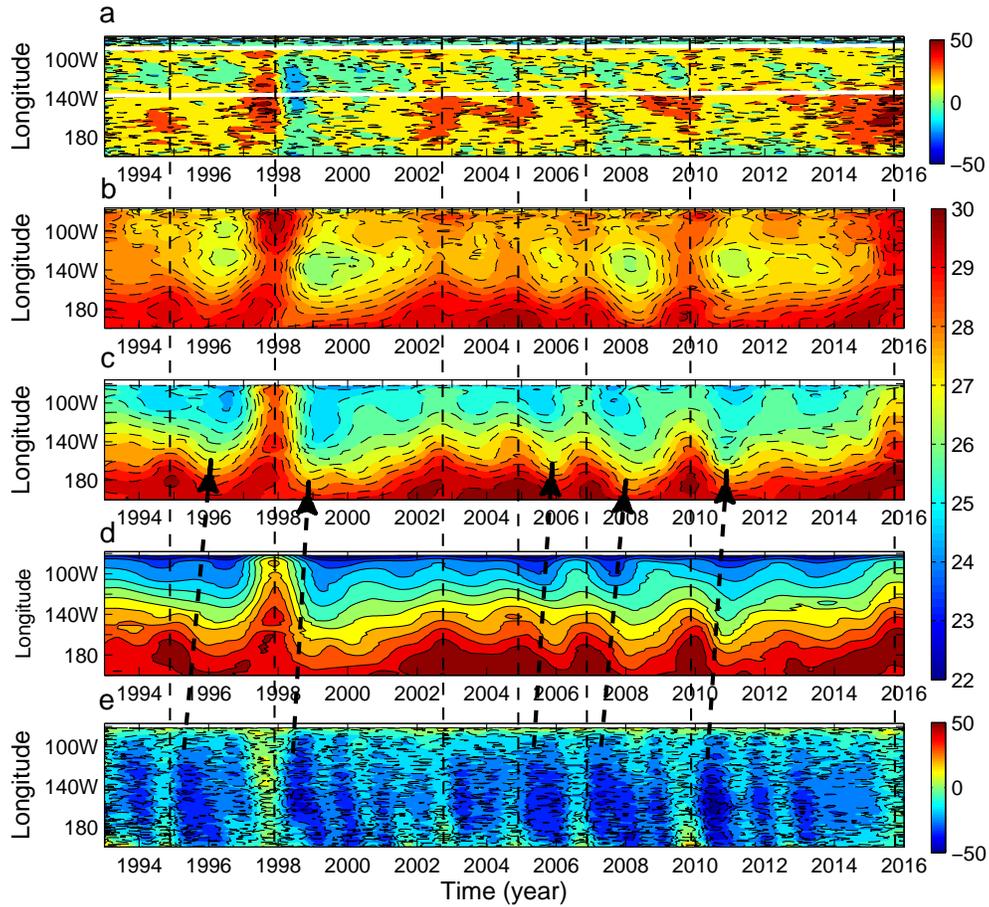}}

\caption{Comparisons of the evolution of zonal currents and SSTs with respect to time and longitude.  \textsf{a} and \textsf{e}: The evolution of the NECC and SSEC averaged over $5^{\circ}$N--$7^{\circ}$N and $5^{\circ}$S--$7^{\circ}$S, respectively. Here only the zonal velocities are concerned. The units of the coloured bar is cm/s (positive to the east). \textsf{b--d}: The evolution of SSTs averaged over $4^{\circ}$N--$8^{\circ}$N, $4^{\circ}$S--$4^{\circ}$N  and $4^{\circ}$S--$8^{\circ}$S, respectively. The units for the coloured bar is $^{\circ}$C.
 In \textsf{a}, the pair of white horizontal solid lines denotes the location of the east low-speed gap of the NECC. The thin, vertical dashed lines denote the prosperous periods for seven El Ni\~{n}o events. The five big arrows denote the corresponding relationships between the strongest cases of the SSEC and the lowest cases of the SST in the CP.
 }
\end{figure}

\begin{figure}[h]
\centerline{\includegraphics[width=\textwidth]{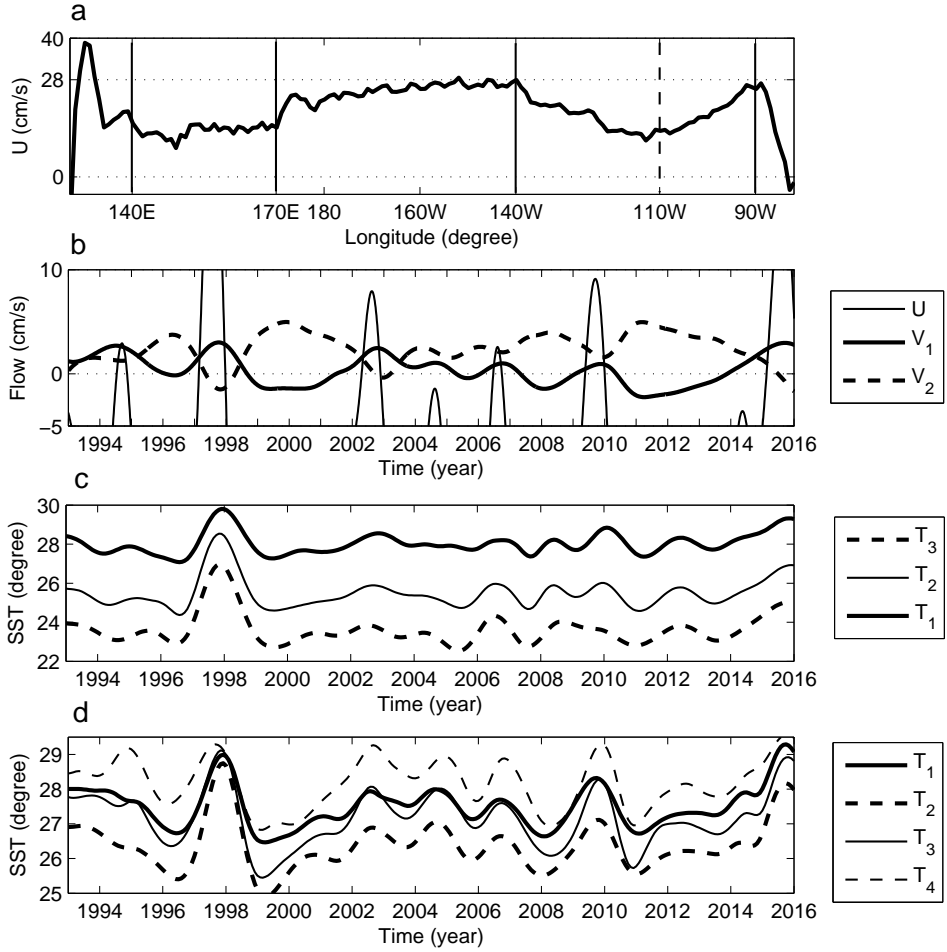}}

\caption{\textsf{a}: Distribution of time-averaged zonal velocity in \textsf{5a} along the longitude. The two pairs of solid vertical lines denote the locations of the western and eastern low-speed gaps of the NECC.
The dashed line at $110^{\circ}$W almost accords with the trough of the east gap. \textsf{b}: Time variations of the mean flows on $170^{\circ}$E--$170^{\circ}$W. U is the zonal velocity of the CECC averaged over $2^{\circ}$S--$2^{\circ}$N (positive is to the east). ${V}_1$ and ${V}_2$ are meridional flows averaged over $3^{\circ}$N--$5^{\circ}$N
and $3^{\circ}$S--$5^{\circ}$S, respectively (positive is to the north).
\textsf{c}: Time variations of the mean SSTs on $90^{\circ}$W--$82^{\circ}$W.  ${T}_1,\; {T}_2$ and ${T}_3$ are the SSTs averaged over $4^{\circ}$N--$8^{\circ}$N, $4^{\circ}$S--$4^{\circ}$N
and $4^{\circ}$S--$8^{\circ}$S, respectively. \textsf{d}: ${T}_1$ stands for the time variation of the SST averaged on the region,
 delimited by $4^{\circ}$N--$8^{\circ}$N and $140^{\circ}$W--$110^{\circ}$W.
${T}_2,\;{T}_3$ and ${T}_4$ stand for the time variation of the SSTs averaged on the regions delimited by $4^{\circ}$S--$4^{\circ}$N and $140^{\circ}$W--$110^{\circ}$W, $150^{\circ}$W--$140^{\circ}$W and $180^{\circ}$W--$150^{\circ}$W, respectively.}
\end{figure}

\begin{figure}[h]
\centerline{\includegraphics[width=\textwidth]{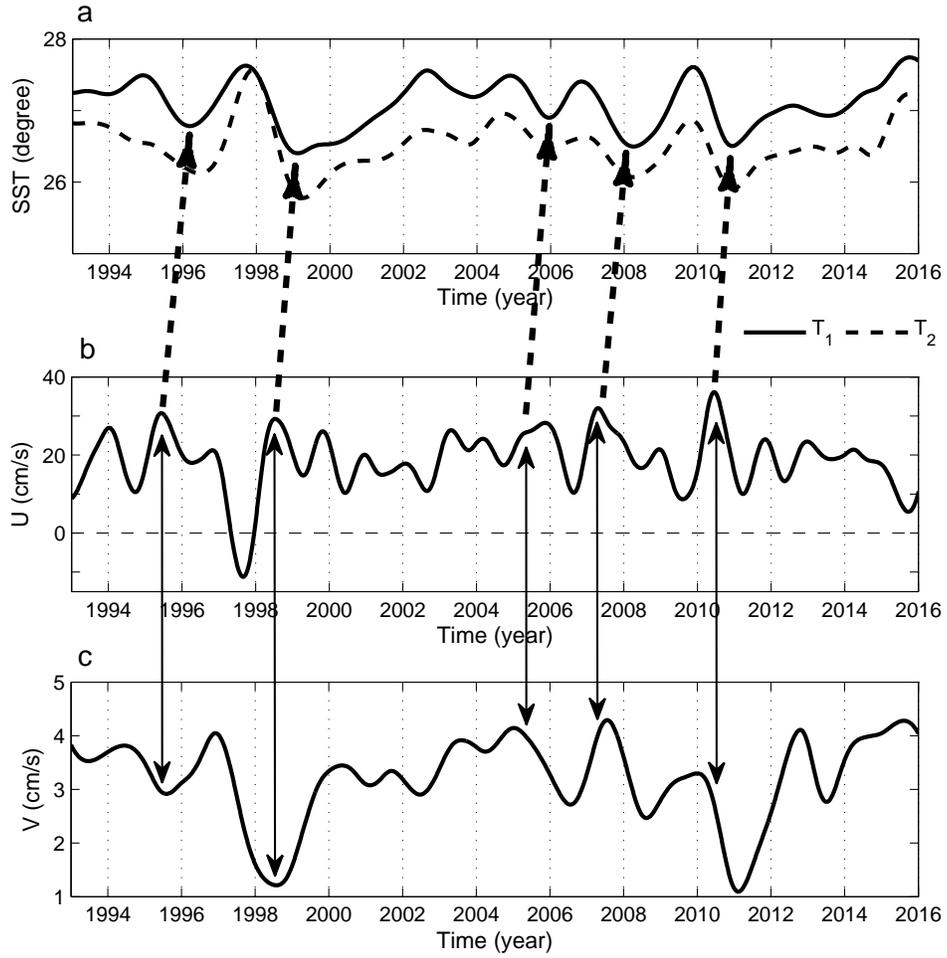}}

\caption{Time variation of the mean temperature and velocity in Ni\~{n}a-A region. \textsf{a}: Comparison between the SST in Ni\~{n}a-A and that in central equatorial Pacific. ${T}_2$ and ${T}_1$ are the SSTs averaged on  Ni\~{n}a-A and on the region bounded by
$4^{\circ}$S--$4^{\circ}$N from $170^{\circ}$E to $140^{\circ}$W, respectively. \textsf{b} and \textsf{c} show the mean zonal (positive is to the west) and meridional (positive is to the north) components of the velocity in Ni\~{n}a-A. The big arrows denote the corresponding relationships between the five lowest cases of SST in the central equatorial Pacific, and the strongest cases of zonal flow of Ni\~{n}a-A.
In addition, the five double-sided arrows further transfer these correspondence to the meridional flow.
 }
\end{figure}

\begin{figure}[h]
\centerline{\includegraphics[width=\textwidth]{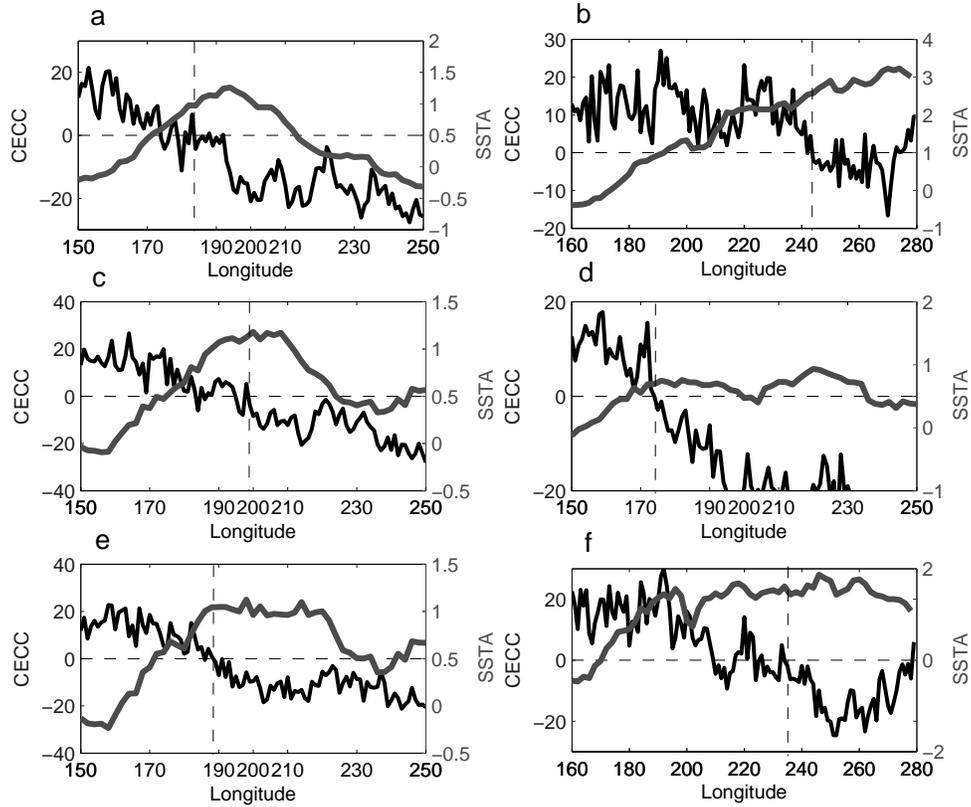}}

\caption{Comparisons between the velocities of the CECC (averaged over $2^{\circ}$S--$2^{\circ}$N, black lines) and the SST anomalies (averaged over $4^{\circ}$S--$4^{\circ}$N, grey lines) with respect to different longitudes and different prosperous periods of El Ni\~{n}o events. From \textsf{a} to \textsf{f}, the corresponding periods are
Nov.1994, Nov.1997, Dec.2002, Oct.2004, Nov.2009 and Oct.2015, respectively.  The vertical dashed line in each sub-figure denotes the easternmost location reached by the CECC. The units for the velocity and temperature are cm/s (positive is to the east) and $^{\circ}$C, respectively. }
\end{figure}

\begin{figure}[h]
\centerline{\includegraphics[width=\textwidth]{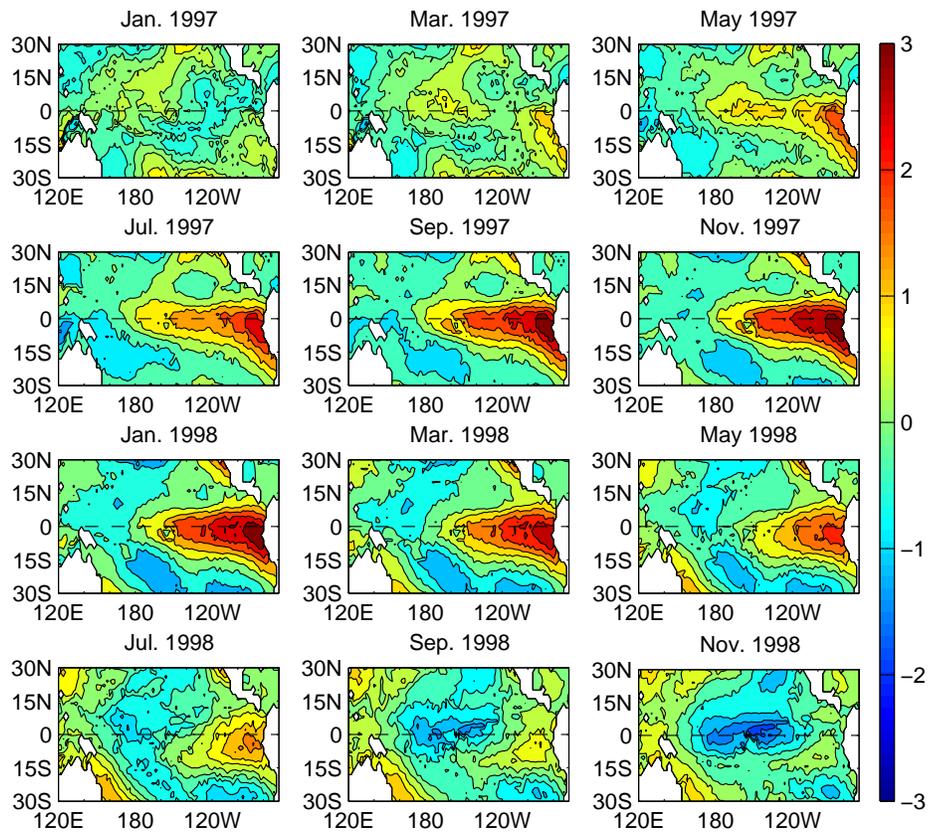}}

\caption{Space-time evolution of the SST anomalies from Jan.1997 to Nov.1998, in an interannual time scale. The units of the coloured bar is $^{\circ}$C. }
\end{figure}

\begin{figure}[h]
\centerline{\includegraphics[width=\textwidth]{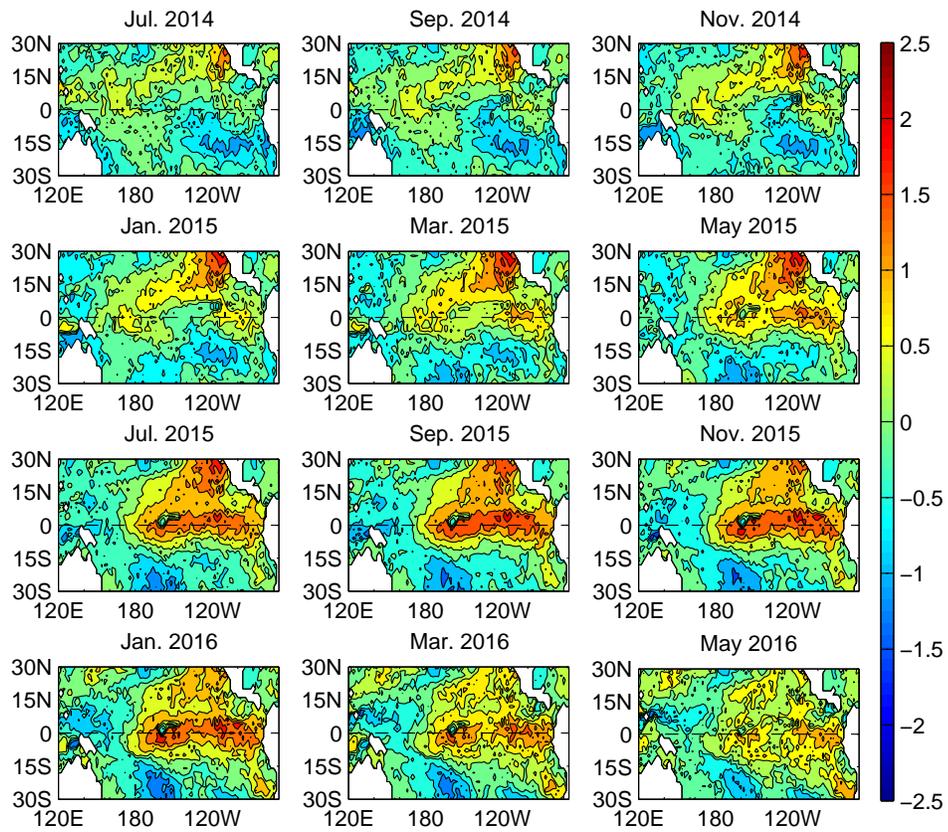}}

\caption{Space-time evolution of the SST anomalies from Jul.2014 to May 2016, in an interannual time scale. The units of the coloured bar is $^{\circ}$C. }
\end{figure}

\end{document}